\documentstyle{aipproc}
\renewcommand{\thefootnote}{\fnsymbol{footnote}}

\begin{document}

\title{The Heavy Quark Parton Oxymoron \\
{\large -- A mini-review of Heavy Quark Production theory in {\normalsize %
PQCD}}\thanks{%
Invited Talk given at DIS97 Workshop/Symposium, Chicago, April 1997. 
The general-mass variable-flavor-number scheme and its applications reported
in this talk are based on work done in collaboration with M. Aivazis, J.
Collins, F. Olness, H. Lai, R. Scalise, J. Amundson, C. Schmidt, and X. Wang.%
}}
MSU-HEP-70624 \hfill CTEQ-724.
\author{Wu-Ki Tung}
\address{Department of Physics and Astronomy, Michigan State University
\\ E. Lansing, Michigan 48824, USA}

\maketitle

\begin{abstract}
Conventional perturbative QCD calculations on the production of a heavy
quark ``$H$'' consist of two contrasting approaches: the usual QCD parton
formalism uses the {\em zero-mass approximation} ($m_H=0$) once above
threshold, and treats $H$ just like the other light partons; on the other
hand, most recent ``NLO'' heavy quark calculations treat $m_H$ as a {\em %
large parameter} and always consider $H$ as a {\em heavy particle}, never as
a parton, irrespective of the energy scale of the physical process. By their
very nature, both these approaches are limited in their regions of
applicability. This dichotomy can be resolved in a unified general-mass
variable-flavor-number scheme, which retains the $m_H$ dependence at all
energies, and which naturally reduces to the two conventional approaches in
their respective region of validity. Recent applications to lepto- and
hadro-production of heavy quarks are briefly summarized.
\end{abstract}

\renewcommand{\thefootnote}{\arabic{footnote}} \setcounter{footnote}{0}

\subsubsection{Introduction}

\label{sec:Intro}

The production of heavy quarks in photo-, lepto-, and hadro-production
processes has become an increasingly important subject of study both
theoretically and experimentally. For a comprehensive review and references,
see Ref. \cite{FMNR97a}. The theory of heavy quark production in
perturbative Quantum Chromodynamics (PQCD) is considerably more subtle than
that of light parton (jet) production because of the additional scale
introduced by the quark mass. Let us consider the production of a generic
heavy quark, denoted by $H$, with non-zero mass $m_H,$ in high energy
interactions. For definiteness and simplicity, unless otherwise stated, we
shall use deep inelastic lepton-hadron scattering as the talking example. A
reasonable criterion for a quark to be called ``heavy'' is $m_H\gg \Lambda
_{QCD}$, so that perturbative QCD is applicable at the scale $m_H$. Thus,
conventionally, \{$c,b,t$\} quarks are regarded as heavy.

\vspace{5pt}The relevant {\bf energy scales of this problem} are: (i) a
typical {\em small scale} such as $\Lambda _{QCD}$ or masses of light
mesons, nucleons, ...; (ii) the {\em highest energy scale} $E$ or $\sqrt{s};$
(iii) a{\em \ typical large scale} in the physical process, such as $p_t$ of
the heavy quark (or the associated heavy flavor hadron), $Q$ of deep
inelastic scattering or Drell-Yan processes, or some large mass (such as $%
\,m_{W,\,Z,\,Higgs,\,SUSY})$ -- to be denoted henceforth collectively as $Q$%
; and (iv) the heavy quark mass $m_H$. By definition, $m_H\gg \Lambda _{QCD};
$ and we need $\sqrt{s}\ $to be fairly large compared to $m_H$ for the
production cross-section to be substantial. Thus the important ratio of
scales remaining which determines the physics of the heavy quark production
process is that between $m_H$ and $Q$.\footnote{%
In this talk, we shall not consider ``small-$x$'' problems associated with
logarithms of the large ratio $\sqrt{s}/Q,$ cf 
\cite{Smallx}.} We shall be mainly concerned with $c$ and $b$ quarks for
which this ratio can vary over a wide range in practice.

\subsubsection{Two Contrasting Conventional Approaches}

\label{sec:ConvSch}

The two conventional approaches to heavy quark production in PQCD can be
summarized by the following contrasting master equations used in the
calculation\footnote{%
If a final state particle $C$ is observed, the factorization formula should
also contain a fragmentation function $d^C(z,\mu )$. We leave out $d^C(z,\mu
)$ here only for simplicity of discussion. All statements concerning the
parton distributions also apply to the fragmentation functions, if present.}
\begin{equation}
\text{ZM\thinspace VFN:\ \ \ }\sigma _{lA\rightarrow CX}=\sum_{a%
\mbox{\scriptsize \ = all active
partons}}f_A^a(x_a,\mu )\otimes \hat{\sigma}_{la\rightarrow CX}(\hat{s}%
,Q,\mu )\mid _{_{m_a=0}}^{\overline{MS}}  \label{ZMVFN}
\end{equation}
\begin{equation}
\text{FFN:\ \ \ \ \ \ }\sigma _{lA\rightarrow HX}=\sum_{a%
\mbox{\scriptsize \ = light
partons only}}f_A^a(x_a,\mu )\otimes \hat{\sigma}_{la\rightarrow HX}^{FFN}(%
\hat{s},Q,m_H,\mu )  \label{FFN}
\end{equation}

\noindent
The {\bf Zero-mass Variable-flavor-number (ZM-VFN) scheme} formula, Eq.\ref
{ZMVFN}, is used routinely in most high energy calculations: in global
analyses of parton distributions, from EHLQ \cite{EHLQ} to MRS\cite{MRS} and
CTEQ\cite{CTEQ}, as well as in all analytic or Monte Carlo programs for
generating SM and new physics cross-sections. In this equation, the parton
label $a$ is summed over all possible {\em active parton species}; $\mu $ is
the factorization and renormalization scale; and $\hat{\sigma}%
_{la\rightarrow CX}$ is the perturbatively calculable hard cross section
involving partons only. ``Active'' partons include all quanta which can
participate effectively in the dynamics at the relevant energy scale $\mu
\,(\sim Q)$ \cite{EHLQ,marciano,ColTun}, {\em including charm and bottom
quarks} at current collider energies. Thus, the active flavor number $n_f$
depends on the energy scale (``resolving power'') of the problem; it is not
fixed at any particular value. The hard cross-section $\hat{\sigma}%
_{la\rightarrow CX}(\hat{s},Q,\mu )$ is calculated in the limit of zero mass
for all the partons, and it is made infra-red safe by dimensional
regularization in the \mbox{\small {$\overline {\rm MS}$}}\ scheme -- hence
the name {\em Zero-mass Variable-flavor-number} (ZM-VFN) scheme.

The advantage of the ZM-VFN scheme is that it is quite simple to implement.
For the light partons $a=$\{$g,u,d,s$\}, $m_a\rightarrow 0$ is a valid
approximation for all hard scale $Q$ (since, by definition, $Q\gg m_a$). But
for a heavy quark $H,$ it is a reasonable approximation only in the high
energy regime $\mu \sim Q\gg m_H;$ and it clearly becomes unreliable in the
intermediate region $Q\sim {\cal O}(m_H).\;$

\vspace{1ex} In contrast to the above, the {\bf fixed-flavor-number (FFN)
scheme}, Eq.\ref{FFN} has been used in most recent fixed-order perturbative
calculations of heavy quark production \cite{GHRS,NDE,Smithetal}. In this
scheme, by definition, only light partons (e.g. $u,d,s$ and $g$ for charm
production) are included in the initial state: the number of parton flavors $%
n_f$ is kept at a fixed value regardless of the energy scales involved ($%
n_f=3,4$ for $c,b$ production respectively). The main feature here is: $H$
is pictured as a {\em heavy particle} -- much in the same way as $W,\;Z,$
and other new heavy particles, and very different from the zero-mass light
partons -- hence the mass $m_H$ is kept exactly in the hard cross-section $%
\hat{\sigma}_{ab\rightarrow HX}^{FFN}(\hat{s},Q,m_H,\mu )$. Typically, the
perturbative $\hat{\sigma}_{ab\rightarrow HX}^{FFN}(\hat{s},Q,m_H,\mu )$
will contain logarithm factors of the form $\alpha _s^n(\mu )\ln
^{n-k-m}(Q/m_H)\ln ^m(\hat{s}/Q^2).$ If $Q\sim m_H$ (and $x\sim Q^2/s$ is
not too small), these factors are under control; and we have effectively a
{\em one large scale} hard process. Hence, the FFN scheme is the natural
scheme to use in the energy region $Q\sim m_H$ -- this is precisely where
the ZM VFN scheme is expected to be inappropriate.

From the heavy particle perspective, this approach also has the advantage of
being conceptually simple, even if the NLO calculation requires considerable
amount of work. However, the sharp distinction drawn between the $H$ quark
and the other light quarks, say between $c$ and $s,$ in this formalism
appears quite unnatural as the hadron system is probed at the scale $\mu
\sim Q$ available in current high energy processes. And it has been known
since the next-to-leading order (NLO) calculations in the FFN scheme were
completed \cite{NDE,Smithetal} that, for both charm and bottom production,
there are two disconcerting features about the results: (i) the NLO
corrections turn out to be of the same numerical magnitude as (in fact,
generally larger than) the leading order (LO) result; and (ii) the
uncertainty of the theoretical calculation, as measured by the dependence of
the calculated cross section on the unphysical scale parameter $\mu $, is as
large in NLO as in LO -- contrary to what is expected from a good
perturbation expansion \cite{ADM90}. These features mean that the truncated
perturbative series in this scheme has left out important physics effects.
Experimentally, it is also known that the measured charm and bottom
production cross sections do not agree with the NLO theoretical predictions,
at least in the overall normalization, even when the scale $\mu $ is allowed
to vary within a reasonable range \cite{ExpXsec}.

This situation may not be all that surprising: for $c$ and $b$ quarks, the
condition $Q\sim m_H$ is not well satisfied in most practical cases. In
fact, current experimental ranges for lepto- and hadro-production of these
heavy flavors mostly lie in a region between those appropriate for the ZM
VFN ( $Q\gg m_H$) and FFN ($Q\sim m_H$) schemes. We need a well-defined
theory which applies over the full $Q$ range! Other possible sources for
these problems are: (i) large corrections due to large logarithms of $\left(
s/Q^2\right) $---the {\em small-x} problem \cite{Smallx}; (ii) inadequate
understanding of the hadronization of heavy quarks in comparing PQCD
calculations with experiment; and (iii) existence of non-perturbative
components of $H$ inside the nucleon which are, by definition, excluded by
the FFN scheme. In this talk, we shall concentrate on physics issues
pertaining to the changing role of the heavy quark $H$ over the full $Q$
range. It is particularly interesting because the interplay between the two
independent scales $m_H$ and $Q$ embodies much interesting QCD physics which
is amenable to precise treatment.

\subsubsection{A Unified, General-mass, Parton Approach}

\label{sec:ACOT}

When the energy scale becomes large, $Q/m_H\gg 1$, the FFN scheme becomes
suspect because large logarithm factors of $\ln (Q/m_H)$ in the hard cross
section $\hat{\sigma}_{ab\rightarrow HX}^{FFN}$ becomes increasingly
singular, and higher-order terms containing higher powers of the same can no
longer be omitted. In other words, the truncated perturbation series in this
scheme can become rather unreliable as $Q$ becomes large. The clue for
addressing this problem is already contained in Eq.\ref{ZMVFN}: large
logarithms of the form $\alpha _s^n\ln ^{n-k}(Q/m_H)$ in $\hat{\sigma}%
_{ab\rightarrow HX}^{FFN}$ can be resummed {\em to all orders} in $\alpha _s$
to become the parton distribution $f_A^H(x,\mu )$ (evolved to ($k+1$%
)-loops). The $H$ parton should be included in the sum over parton flavors;
it participates in the hard scattering on the same footing as the other
partons. After removing these potentially dangerous logarithm terms, the
remaining hard cross section $\hat{\sigma}_{ab\rightarrow HX}$ becomes
infra-red safe as $Q/m_H\rightarrow \infty $. It is important to note,
however, the resummation of large $\ln (Q/m_H)$ logarithms does not require
taking the $m_H\rightarrow 0$ limit for the remaining (``mass-subtracted'')
hard cross-section as is done in the conventional ZM VFN formulation, Eq.\ref
{ZMVFN}. In fact, by retaining the $m_H$ dependence in the mass-subtracted
(hence infra-red safe) $\hat{\sigma}_{ab\rightarrow HX}(\hat{s},Q,m_H,\mu ),$
one arrives at a consistent theory for heavy quark production which is valid
over the entire energy range from $Q\lesssim m_H\ \ $to $Q\gg m_H$:\footnote{%
Factorization of any applicable fragmentation functions is implicitly
assumed.}
\begin{equation}
\sigma _{lA\rightarrow CX}(s,Q,m_H)=\sum_{a%
\mbox{\scriptsize \ = all active
partons}}f_A^a(x_a,\mu )\otimes \hat{\sigma}_{la\rightarrow CX}(\hat{s}%
,Q,m_H,\mu )  \label{GMfacthm}
\end{equation}
A program to systematically implement this intuitive physical picture has
been developed in a series of papers in \cite{ColTun,OlnTun,AOT,ACOT}. The
resulting formalism constitutes a natural generalization of the conventional
zero-mass QCD parton formalism to correctly include general quark mass
effects, hence will be called the general-mass variable-flavor-number
(GM-VFN) scheme. (In some recent literature it has also been called the ACOT
scheme, Ref. \cite{ACOT}.)

More precisely, this formalism is based on a well defined renormalization
scheme \cite{CWZ} which provides a natural transition from the threshold
region $Q\sim {\cal O}(m_H)$ to the high energy region $Q\gg m_H$; and the
validity of the generalized factorization theorem can be established
order-by-order in perturbation theory \cite{Collins97}. The key points are:

\begin{itemize}
\item  the renormalization scheme is a composite of two simple schemes,
natural for $Q\lesssim m_H$ and $Q\geq m_H$ respectively, with matching
conditions that make the schemes equivalent in the domain of overlap $Q\sim
m_H$ where they are equally valid for practical low order calculations \cite
{CWZ};

\item  one scheme utilizes a subtraction procedure (BPHZ) which leads to
manifest decoupling of the heavy particle in the region $Q\ll m_H$, thereby
gives precise meaning to the FFN scheme (with no heavy quark partons);

\item  the other scheme is ordinary \mbox{\small {$\overline {\rm MS}$}}\ as
regards the definition of the coupling $\alpha _s(\mu )$ and the parton
densities $f_A^a(x,\mu )$, hence retains the normal $\left( m_H=0\right) $
evolution equations for the latter in the region $Q\gtrsim m_H$ \cite{ColTun}
-- this comes about because the evolution kernels are anomalous dimensions
which are derivatives of renormalization constants, and renormalization
constants are mass-independent in the \mbox{\small {$\overline {\rm MS}$}}\
scheme;

\item  the factorization scheme is defined such that all infra-red safe $m_H$%
-dependent effects are preserved in the hard cross-sections, so that there
is no loss of accuracy when $Q\sim m_H$ \cite{ACOT}. This is accomplished by
defining $\hat{\sigma}_{la\rightarrow CX}(\hat{s},Q,m_H,\mu )$ as the full $%
\sigma _{la\rightarrow CX}(\hat{s},Q,m_H,\mu )$\footnote{%
Needless to say, colinear singularities associated with light partons are %
\mbox{\small {$\overline {\rm MS}$}}\ subtracted.} with mass ($m_H$)
singularities subtracted.
\end{itemize}

These features guarantee that predictions of this formalism: (a) coincide
with those of the FFN scheme in its region of applicability, $Q\lesssim m_H;$
(b) reduce to those of the conventional zero-mass parton model in the
asymptotic energy regime $Q\gg m_H;$ and (c) provide a good approximation to
the physical cross-section over the entire energy range in between, since
the remainder of the perturbation series contains no large logarithms of the
kind $log(Q/m_H).$\footnote{%
As mentioned earlier, we do not consider ``small-$x$'' corrections in this
paper.}

\subsubsection{How, and Why, does the ACOT scheme work?}

\label{sec:HowWhy}How and why does this scheme work were described in some
detail for lepto-production of heavy quarks in Ref.~\cite{ACOT}. We recall
the essential points here. Writing out the first two terms in the
perturbative expansion of Eq.\ref{GMfacthm}, we have
\begin{equation}
\sigma _{lA\rightarrow HX}(s,Q,m_H)=f_A^H\,\otimes \,^0\hat{\sigma}%
_{lH\rightarrow lH}\,+\,f_A^g\,\otimes \,^1\hat{\sigma}_{l\,g\rightarrow lH%
\bar{H}}  \label{HadFact1}
\end{equation}
where the superscript $n$ in $^n\hat{\sigma}$ denotes the order in $\alpha
_s $ of the hard cross-section $\hat{\sigma}$. The lowest order hard
cross-section $^0\hat{\sigma}_{lH\rightarrow lH}$ is identical to the Born
expression $^0\sigma _{lH\rightarrow lH}$ since the tree diagram does not
need any subtraction. The order $\alpha_s$ hard cross-section is given by
\begin{equation}
\,^1\hat{\sigma}_{l\,g\rightarrow lH\bar{H}}=\,^1\sigma _{l\,g\rightarrow lH%
\bar{H}}\,-\,^1\,f_g^H\,\otimes \,^0\hat{\sigma}_{lH\rightarrow
lH}\;;\;\,^1\,f_g^H\,=\frac{\alpha _s(\mu )}{2\pi }\ln \frac{\mu ^2}{m_H^2}%
P_{gH}\,  \label{ParEq}
\end{equation}
where $^1\,f_g^H$ is the perturbative parton distribution function of
finding $H$ in $g;$ $P_{gH}$ is the usual $g\rightarrow H$ splitting
function; and $^1\sigma _{l\,g\rightarrow lH\bar{H}}$ is the {\em %
unsubtracted cross-section}\footnote{%
Here ``unsubtracted'' refers to heavy quark mass effects. As mentioned
before, colinear singularities due to light partons are always subtracted as
in the \mbox{\small {$\overline {\rm MS}$}}\ scheme.} for the gluon fusion
process $l\,g\rightarrow lH\bar{H}$. The subtraction term above can be
formally derived by applying Eq.\ref{HadFact1} to the partonic cross-section
$^1\sigma_{l\,g\rightarrow lH\bar{H}}$ (with light-parton colinear
singularities subtracted), and then solving for $^1\hat{\sigma}%
_{l\,g\rightarrow lH\bar{H}}$ \cite{ACOT}.

Substituting Eq.\ref{ParEq} in Eq.\ref{HadFact1}, the physical cross-section
becomes:
\begin{equation}
\begin{array}{ccl}
\sigma _{lA\rightarrow HX} & = & f_A^H\,\otimes \,^0\sigma _{lH\rightarrow
lH}\,+\,f_A^g\,\otimes (\,^1\sigma _{l\,g\rightarrow lH\bar{H}%
}\,-\,^1\,f_g^H\,\otimes \,^0\sigma _{lH\rightarrow lH}) \\ \rule{0mm}{3.2ex}
& = & (f_A^H\,\,-\,f_A^g\,\otimes \,^1\,f_g^H\,)\otimes \,^0\sigma
_{lH\rightarrow lH}+\,f_A^g\,\otimes \,^1\sigma _{l\,g\rightarrow lH\bar{H}}
\end{array}
\label{HadFact2}
\end{equation}
The three terms on the right-hand-side of the first line are: {\em %
heavy-flavor excitation} (HE), {\em heavy-flavor creation} (HC), and the
subtraction term; the last physically represents the overlap between the two
production mechanisms. Both $^0\sigma _{lH\rightarrow lH}\,$ and $^1\sigma
_{l\,g\rightarrow lH\bar{H}}$ (unsubtracted) contain the full $m_H$
dependence. For $Q\gg m_H,$ it is useful to view the right-hand side of Eq.%
\ref{HadFact2} as on the first line. The quantity in the parenthesis (really
just $\,^1\hat{\sigma}_{l\,g\rightarrow lH\bar{H}}$) is free of large $\ln
\frac{Q^2}{m_H^2}$ logarithms because of the subtraction; it is infra-red
safe. The whole term remains at numerical order $\alpha _s\times {\cal O}(1)$
(in contrast to the unsubtracted $^1\sigma _{l\,g\rightarrow lH\bar{H}}$
which is the LO FFN scheme result with a large log factor,
$\alpha _s\times {\cal O}(\ln \frac{Q^2}{m_H^2})$) in the large $Q$ limit.
Thus, one recovers the LO ZM VFS formula with the HE contribution as the
dominant term for the cross-section. In the threshold region, $Q\sim m_H,$
it is more useful to focus attention on the second line in Eq.\ref{HadFact2}
: the two terms inside the parenthesis are individually small in this region
and, in addition, they cancel each other up to order $\alpha _s^2$ since
both satisfy the evolution equation (to this order) with the same boundary
conditions (assuming there is no non-perturbative charm). As a consequence,
the full cross-section is dominated by the second (HC) term -- which is the
FFN scheme result to order $\alpha _s$.

The explicit expression for the subtraction term clearly shows how it
overlaps the HE and HC mechanisms, hence leads to the appropriate
cancellations in the respective kinematic regions. The same principle
applies in other lepton-hadron and hadron-hadron processes to all orders of
perturbation theory \cite{Collins97}. Numerical calculations based on Eq.\ref
{HadFact2} confirm the features described above for relevant physical
cross-sections. See Ref.\cite{ACOT} and the talk by Schmidt \cite
{SchmedtDis97} for plots of $F_2^c\left( x,Q\right) $ with individual
contributions from the three terms on the right-hand side of Eq.\ref
{HadFact2} which explicitly illustrate these features.

\subsubsection{Complementarity between HE and High-order HC Contributions}

\label{sec:Complement}

The GM-VFN (ACOT) scheme highlights some overlapping features of HE and
higher order HC mechanisms for heavy quark production (hence the need for
the subtraction to avoid double counting): the two are not mutually
exclusive, as sometimes perceived; rather, {\em they are complementary}. For
the total inclusive cross-section (e.g. structure functions $F_{2,3}^c\left(
x,Q\right) $ in DIS), in particular, the HE contribution represents the
result of resumming the colinear parts of HC diagrams to all orders in the
running coupling. This provides an efficient method of obtaining important
quantitative results without having to calculate many complicated higher
order diagrams in the FFN scheme.

The trade-off is that some information on the differential distributions
(e.g. transverse momentum spectrum of the heavy particle) is integrated over
in the resummation, hence becomes less accessible. For example, if one is
interested in the $p_t$ distribution of the charm quark in DIS
lepto-production with respect to the virtual photon-target axis, the HE
contribution, as well as the subtraction term, in Eq.\ref{HadFact2} will be
formally proportional to the delta function $\delta (p_t)$. This is,
mathematically, a {\em distribution} (rather than an ordinary function)
which needs to be folded with a suitable {\em smearing function }-- some
combination of experimental resolution function and theoretical $p_t$
distribution (see below) -- in order to produce meaningful physical
cross-sections. Away from the small $p_t\ $region, the low-order FFN scheme
diagrams will be the logical place to start in obtaining the leading
contributions to the physical $p_t$ spectrum. Cf. the talk by Schmidt \cite
{SchmedtDis97}. To obtain a precise theory of $p_t$ distributions over the
entire range, a different kind of resummation will be required. Thus, for
heavy quark production in PQCD, as in many quantum mechanical problems in
general (say, the double-slit problem), the appropriate way to formulate the
theory depends intimately on the physics question asked.

\subsubsection{Recent Developments}

\label{sec:NewDev}

The ACOT scheme has been applied to the analysis of recent charm production
data in neutrino scattering \cite{CCFRcharm}, yielding the most up-to-date
information on the strange quark content of the nucleon.

Recently, in the wake of new precision measurements of the total $F_2\left(
x,Q\right) $ at small $x$ \cite{HERA97} (where charm final states consist of
up to 25\% of the cross-section) as well as first measurements of $%
F_2^c\left( x,Q\right) $ \cite{HERAcharm}, it has become obvious that a more
precise theoretical treatment of charm production in NC DIS than those used
so far (the ZM-VFN and FFN schemes described earlier) is now needed. A first
step is taken by the CTEQ group, performing a new global QCD analysis of
parton distributions \cite{LaiTun97a} based on the GM-VFN scheme of ACOT (to
order $\alpha _s$ only). A similar study has been done by MRRS \cite{MRRS},
using a related procedure (see also \cite{Roberts97}). An order $\alpha _s^2$
calculation in the general mass scheme is within reach \cite{BuzaEtal} since
much of the known FFN scheme results to this order can be used as the
starting point.

Of equal importance is an improved theoretical treatment of various heavy
quark cross-sections in hadron-hadron scattering beyond the conventional
calculational schemes, particularly the NLO FFN scheme results which suffer
from the problems mentioned in Sec.\ref{sec:ConvSch}. Some preliminary
results have been obtained in the ACOT scheme \cite{OST96,COST97}. The
qualitative features of these results, compared to existing calculations,
are similar to those on lepto-production described briefly in Sec.\ref
{sec:HowWhy}. As expected, the inclusion of heavy-flavor-excitation
contributions leads to: (i) reduced dependence on the (unphysical) scale
parameter $\mu ;$ and (ii) an increase in the predicted cross-section over
most of the range of $p_t$ of the produced heavy quark, as preferred by data
\cite{ExpXsec}. Detailed phenomenology of the inclusive cross-section, as
well as correlations of final-state heavy particles still need to be pursued
\cite{COST97}.

\subsubsection{Concluding Remarks}

The general-mass variable-flavor-number scheme of Ref.~\cite
{ColTun,AOT,ACOT,Collins97} generalizes the familiar zero-mass
variable-flavor-number QCD parton framework (valid only at large $Q$ scales)
to include quark mass effects, and it reproduces the results of the widely
used FFN scheme heavy quark calculations (valid in the one-large-scale
region $Q$ $\sim $ $m_H$). It naturally unifies the two contrasting
conventional approaches in a well-defined renormalization and factorization
scheme. First results on lepto- and hadro-production demonstrate
improvements over existing calculations both in smaller $\mu $-dependence
and in increased cross-sections.

However, the more complete theory is not a cure-for-all. Although detailed
phenomenology has yet to be done, there is no doubt that further
developments in several directions are needed to reach a full understanding
of heavy quark production in QCD. For instance, (i) the general mass scheme
should be implemented to higher orders for both lepto- and hadro-production;
(ii) the significance of possible large logarithms of the type $\ln
(Q^2/s),\ln (m_H^2/s)$ $\sim \ln x$ -- the {\em small-x problem} -- needs to
be better understood \cite{Smallx}; (iii) both perturbative and
non-perturbative aspects of the hadronization of the heavy quarks deserve
further study; and (iv) the question ``Are there non-perturbative
charm/bottom components inside hadrons?'' needs to be answered. The last
question, having been discussed in the literature for many years, has to be
carefully investigated phenomenologically; and this can be done only in the
framework of the general-mass scheme, since the existence of
non-perturbative non-zero-mass parton is excluded by assumption in the FFN
scheme.

\end{document}